\documentclass{PoS}

\usepackage{amssymb,amsmath}

\title{One-loop Amplitudes as BPS state sums}

\ShortTitle{One-loop Amplitudes as BPS state sums}

\author{\speaker{Ioannis Florakis}\thanks{Based on joint work with C. Angelantonj and B. Pioline, both of whom I would like to warmly thank for a very enjoyable collaboration.}\\
        Max-Planck-Institut f\"ur Physik\\
        Werner-Heisenberg-Institut, 
        F\"ohringer Ring 6, 80805 M\"unchen, Germany\\
        E-mail: \email{  florakis@mppmu.mpg.de }}

\abstract{We review a novel method for evaluating one-loop BPS-saturated amplitudes in string theory. Contrary to traditional techniques of unfolding the fundamental domain $\mathcal{F}$ against the Narain lattice, which are only valid in certain regions  of the moduli space and which obscure the T-duality invariance of the result, we will describe how the elliptic genus can be represented as a linear combination of certain absolutely convergent Poincar\'e series, against which $\mathcal{F}$ can be unfolded. The result can be expressed as a sum of one-loop contributions of  perturbative BPS-states in a manifestly T-duality invariant fashion, valid at any point of (the perturbative) moduli space.  Within this framework, the singularity structure of  amplitudes around points of gauge symmetry enhancement becomes crystal clear and a series of applications  is given in order to better illustrate the power of this approach.}

\FullConference{Proceedings of the Corfu Summer Institute 2012 "School and Workshops on Elementary Particle Physics and Gravity"\\
		September 8-27, 2012\\
		Corfu, Greece}

\begin{document}
 
\section{Introduction}

The problem of calculating scattering amplitudes in string theory lies at the core of any attempt to make contact with low-energy phenomenology. Indeed, string phenomenology has been marked with impressive progress during the last two decades and several semi-realistic models have been constructed which, at tree-level, provide viable candidates for the description of supersymmetric extensions of the Standard Model, including particle content and interactions. To this end, the need for incorporating loop corrections to gauge and gravitational couplings in the effective action is inherently linked with any attempt to make further contact with experimental data. Furthermore, and regardless of the possible direct applications to low-energy phenomenology, the development of a powerful framework for the study of stringy corrections to effective couplings acquires a theoretical importance on its own especially since, in the presence of sufficient supersymmetry, several BPS-saturated couplings are protected against higher perturbative or even non-perturbative corrections and provide useful laboratories in which to test string dualities (see e.g.\cite{Kiritsis:1999ss} and references therein).

In closed (oriented) string perturbation theory, one is typically dealing with a topological (Polyakov) expansion over closed, genus-$g$ Riemann surfaces, which we schematically write as:
\begin{align}
	\sum\limits_{g=0}^{\infty}{g_s^{2(g-1)}}\int\limits_{\textrm{moduli}}~\int\limits_{\textrm{vertex operator}\atop\textrm{insertions--}z_i}~\int{\mathcal{D}X\,\mathcal{D}\psi\,\ldots \mathcal{V}_i(z_i)\ldots } ~e^{-S[X,\psi,g_{ab},\ldots]}~,
\end{align}
where $g_s$ is the string coupling and $S[X,\psi,g_{ab},\ldots]$ is the worldsheet action. After appropriately gauge-fixing the diffeomorphism and Weyl gauge symmetries and performing the path integral over the various worldsheet fields $X,\psi$, one is instructed to integrate over the worldsheet positions $z_i$ of the various vertex operator insertions $\mathcal{V}_i(z_i)$ and, eventually, integrate over the moduli space of the Riemann surface with an appropriate measure. We will focus entirely on one-loop amplitudes ($g=1$), for which the latter is an integral over the complex structure $\tau\in\mathcal{H}$ of the worldsheet torus, of the generic form:
\begin{align}\label{genericInt}
	\int\limits_{\mathcal{F}}{d\mu~\mathcal{A}(\tau,\bar\tau)}~.
\end{align}
The Teichm\"uller parameter $\tau$ is initially defined over the upper half-plane $\mathcal{H}$, before gauging the residual discrete group $PSL(2;\mathbb{Z})$ of large diffeomorphisms, also known as the modular group, restricts integration down to its fundamental domain under the modular group, $\mathcal{F}=\mathcal{H}/PSL(2;\mathbb{Z})$. The $SL(2;\mathbb{R})$-invariant measure will, henceforth, be denoted by $d\mu \equiv d^2\tau/\tau_2^2$. The function $\mathcal{A}(\tau,\bar\tau)$, obtained after performing the path integral and integration over the positions of the insertions $\mathcal{V}_i$, is, by consistency, modular invariant. 

The techniques that we will review in this note concern precisely the evaluation of genus-1 modular integrals of the form (\ref{genericInt}), for certain classes of the integrand functions $\mathcal{A}$, which appear naturally in string theory. Contrary to the traditional `orbit method' used in the literature \cite{McClain:1986id,O'Brien:1987pn,Dixon:1990pc,Mayr:1993mq,Bachas:1997mc,stst1,stst2,Kiritsis:1997hf,Kiritsis:1997em,Marino:1998pg,Harvey:1995fq,Harvey:1996gc}, the new methods we present \cite{Angelantonj:2011br,Angelantonj:2012gw} manifestly preserve the T-duality symmetries of the theory, provide a natural modular invariant IR regularization and clearly exhibit the singularity structure of the associated amplitudes, without depending on the region in moduli space around which one is unfolding.
There are four major cases of interest, of increasing difficulty, depending on the holomorphicity of $\mathcal{A}$ :
\begin{align*}
	\begin{array}{c || c || c}
		\textrm{Case} &\mathcal{A} & \textrm{Type}\\  \hline
		\textrm{I} &\Phi(\tau) & \textrm{holomorphic function of}~ \tau\\
		\textrm{II} &\Gamma_{(d,d)}(\tau,\bar\tau) & \textrm{ Narain lattice} \\
		\textrm{III} &~~\Gamma_{(d+k,d)}(\tau,\bar\tau)\Phi(\tau)~~ & ~~~~\textrm{lattice with Wilson lines times elliptic genus}~\Phi\\
		\textrm{IV} &\mathcal{Z}(\tau,\bar\tau) & \textrm{manifestly non-holomorphic function}\\
	\end{array}
\end{align*}
Case I involves a holomorphic modular invariant function of $\tau$ and can be treated in a straightforward fashion using Stokes' theorem \cite{Lerche:1987qk}, in addition to the new methods we will review here. Case II involves the integral of a $d$-dimensional lattice and can be considered a special case of III, where an asymmetric\footnote{The asymmetry $k$ of the lattice can be thought of as parametrizing the presence of non-trivial Wilson lines of a Heterotic compactification.} lattice is multiplied together with an holomorphic function $\Phi(\tau)$ of modular weight $w=-k/2$, known as the (modified) elliptic genus. Case III is the generic form of special classes of $\mathcal{N}=2$ amplitudes, known as BPS-saturated couplings. Due to the presence of extended $\mathcal{N}=2$ supersymmetry, which is responsible for the factorization of $\mathcal{A}$ into a lattice $\Gamma_{(d+k,d)}$ times the elliptic genus $\Phi(\tau)$, perturbative corrections to these couplings may terminate at 1-loop. Typical examples are the $\mathcal{N}=2$ gauge and gravitational threshold corrections (\emph{cf.} \cite{Harvey:1995fq,Harvey:1996gc}) in Heterotic string theory compactified on $K3\times T^2$:
\begin{align}
	\begin{split}
	&\Delta_G = -i\int\limits_{\mathcal{F}}d\mu~\biggr[\frac{\tau_2}{2\eta^2}\,\textrm{Tr}\biggr\{ J_0\,e^{i\pi J_0}\,q^{L_0-\frac{c}{24}}\,\bar{q}^{\bar{L}_0-\frac{\hat{c}}{24}} \left(Q^2-\frac{1}{4\pi\tau_2}\right)\biggr\}-b_G\,\tau_2\biggr]~,\\
	&\Delta_{\textrm{grav}} = -i\int\limits_{\mathcal{F}}d\mu~\biggr[\frac{\tau_2}{2\eta^2}\,\textrm{Tr}\biggr\{ J_0\,e^{i\pi J_0}\,q^{L_0-\frac{c}{24}}\,\bar{q}^{\bar{L}_0-\frac{\hat{c}}{24}} \biggr\}\frac{\hat{E}_2}{12}-b_{\textrm{grav}}\,\tau_2\biggr]~,
	\end{split}
\end{align}
where the traces run over the internal $(9,22)$ superconformal (SCFT) theory, with the right-movers\footnote{For notational simplicity, we adopt the convention where the supersymmetric side of the Heterotic string  is taken to be the right-moving, holomorphic side.} set to their Ramond ground state and $J_0$ is the $U(1)$ generator of the internal $\mathcal{N}=2$ SCFT. 
The trace in the second line is identified with the modified elliptic genus times the $T^2$-lattice and $Q$ denotes one of the Cartan generators of the gauge group factor $G$ whose 1-loop correction is being computed. Furthermore, $b_G,b_\textrm{grav}$ are the coefficients of the associated 1-loop beta functions. The holomorphicity breaking term $1/(4\pi\tau_2)$ in the group trace in $\Delta_G$ arises from a contact term in the Kac-Moody gauge current correlator. Taking the difference of thresholds for two different gauge groups of the same Kac-Moody level (here we consider $k_{G_i}=1$), results in the cancellation of the universal contact terms. In the absence of Wilson lines, the resulting modular integral for the difference $\Delta_{G_1}-\Delta_{G_2}$  will again involve the $T^2$-lattice times a weak holomorphic modular function ($w=0$). Since the unphysical tachyon pole is chargeless under $G_i$, it will not contribute to the difference of quadratic Casimirs $\textrm{Tr}(Q_1^2-Q_2^2)$ and the resulting modular function will be holomorphic everywhere, including the cusp at $\tau=i\infty$. It is then a standard result in the theory of modular forms that an everywhere holomorphic modular function ($w=0$) is actually constant and the difference of thresholds will produce an integral of the $\Gamma_{(2,2)}$ lattice alone. This falls precisely in the class of modular integrals of Case II and is, indeed, encountered \emph{e.g.} when one is calculating the difference between the $E_7$ and $E_8$ group thresholds in $K3\times T^2$ compactifications of the $E_8\times E_8$ Heterotic string. 

%%%%%%

\section{The unfolding method}

For simplicity and concreteness, we will henceforth restrict our attention to the modular integrals of Case III:
\begin{align}\label{GenericBPSIntegral}
	 \int\limits_{\mathcal{F}}{d\mu~\Gamma_{(d+k,d)}(G,B,Y)~\Phi(\tau)} ~,
\end{align}
which appear naturally in 1-loop corrections BPS-saturated couplings in the effective action. The Narain lattice $\Gamma_{(d+k,d)}$  depends on the compactification moduli $G_{IJ},B_{IJ},Y_I^a$ parametrizing the coset $\frac{SO(d+k,d)}{SO(d+k)\times SO(d)}$, with $I=1,\ldots d$ and $a=1,\ldots k$. In the most general case, $\Phi(\tau)$ is a weak almost holomorphic modular form of $SL(2;\mathbb{Z})$, with modular weight $w=-k/2$ and with at most a simple pole at the cusp\footnote{In the case of $SL(2;\mathbb{Z})$, the only cusp is the point at $\tau=i\infty$. }. The latter requirement derives from the reparametrization properties of the string worldsheet. The major difficulty with evaluating (\ref{GenericBPSIntegral}) lies in the non-rectangular shape of the fundamental domain $\mathcal{F}=\{\tau\in\mathcal{H}~:~|\tau|\geq 1~,~|\tau_1|\leq 1/2\}$ and the general technique for evaluating such integrals is  known as the `orbit method'  or the `unfolding method', which we now briefly review. 

Start from the generic integral:
\begin{align}\label{Unfold1}
	I=\int\limits_{\mathcal{F}}d\mu~f(\tau,\bar\tau)~,
\end{align}
where $f(\tau,\bar\tau)$ is a modular function. The idea behind the unfolding method is to exploit modular transformations in order to simplify the integration region, by rewriting (\ref{Unfold1}) as an integral over a rectangular integration domain. It relies on one's ability to express $f$ as a sum over modular orbits or, technically, in finding a Poincar\'e series representation for $f$:
\begin{align}\label{PoincareRep}
	f(\tau,\bar\tau)=\sum\limits_{\gamma\in \Gamma_{\infty}\backslash SL(2;\mathbb{Z})}{ \varphi(\gamma\cdot\tau, \gamma\cdot\bar\tau)}~,
\end{align}
where $\Gamma_\infty$ is the stabilizer of the cusp:
\begin{align}
	\Gamma_\infty  =  \biggr\{ \biggr(\begin{array}{c c} 1 & n \\ 0 & 1\\ \end{array}\biggr)~:~n\in\mathbb{Z}\biggr\} \subset SL(2;\mathbb{Z})~,
\end{align}
and the element $\gamma= \left( {a \atop c}~{b\atop d}\right) \in \Gamma_\infty\backslash SL(2;\mathbb{Z})$ acts on $\tau\in\mathcal{H}$ by linear fractional transformations $\gamma\cdot\tau= \frac{a\tau+b}{c\tau+d}$. The function $\varphi$ in (\ref{PoincareRep}) is known as the `seed', since its modular averaging produces the modular function $f$, and is  assumed to be invariant under rigid translations $\gamma\in \Gamma_\infty$. Using the Poincar\'e series representation (\ref{PoincareRep}) into the integral (\ref{Unfold1}) and, assuming absolute convergence, so that we can pull the sum outside the integral, we obtain after a change of variables $\tau'=\gamma\cdot\tau$:
\begin{align}\label{Unfold2}
	I =\sum\limits_{\gamma\in \Gamma_{\infty}\backslash SL(2;\mathbb{Z})}~ \int\limits_{\mathcal{F}}{ d\mu~\varphi(\gamma\cdot\tau,\gamma\cdot\bar\tau)}=\sum\limits_{\gamma\in \Gamma_{\infty}\backslash SL(2;\mathbb{Z})}~ \int\limits_{\gamma\mathcal{F}}{ d\mu~\varphi(\tau',\bar\tau')} = \int\limits_{\Gamma_\infty\backslash\mathcal{H}}d\mu~\varphi(\tau,\bar\tau)~.
\end{align}
Hence, summing over $SL(2;\mathbb{Z})$ orbits, we see that $\mathcal{F}$ has been `unfolded'  to the half-infinite  strip $S\equiv \Gamma_\infty\backslash\mathcal{H}=\{\tau\in\mathcal{H}~:~-\frac{1}{2}\leq\tau_1<\frac{1}{2}~,~ \tau_2>0\}$. As a result, the original modular integral (\ref{Unfold1}) has been reduced to a simpler integral over the strip, involving the seed $\varphi$ instead of the original function $f$. In fact, the new integration domain being rectangular, the last integral can now be given a `field-theoretic' interpretation, in the sense that we can now consider the $\tau_1$-integral to be imposing level-matching, whereas the $\tau_2$-integral will later turn out to provide a Schwinger-like representation of the amplitude.

%%%%%%

\section{Traditional unfolding of $\mathcal{F}$ against the lattice}\label{sec:TraditionalUnfolding}

So far, the traditional approach in the literature has been to use the orbit decomposition of the Narain lattice in order to evaluate modular integrals of the form (\ref{GenericBPSIntegral}). This has both advantages and disadvantages that we will shortly discuss. We illustrate the unfolding of $\mathcal{F}$ against the lattice with a simple example of this type :
\begin{align}\label{SimpleExample}
	I=\int\limits_{\mathcal{F}}d\mu~ \Gamma_{(1,1)}(R)\,j(\tau)~,
\end{align}
involving a 1-dimensional lattice, corresponding to a circle $S^1$ of radius\footnote{For simplicity, we set everywhere $\alpha'=1$.} $R$:
\begin{align}\label{1dlattice}
	\Gamma_{(1,1)}(R)=R\sum\limits_{\tilde{m},n\in\mathbb{Z}}e^{-\frac{\pi R^2}{\tau_2}|\tilde{m}+\tau n|^2}~,
\end{align}
 times the holomorphic Klein $j$-invariant function. The latter is the unique holomorphic modular function with a first-order pole in the $q=e^{2\pi i\tau}$-expansion at the cusp and is conventionally defined with vanishing constant term, \emph{i.e.} $j(\tau)=\frac{1}{q}+\mathcal{O}(q)$. The partition function of the lattice (\ref{1dlattice}) is given in its Lagrangian representation, with $\tilde{m},n$ being the two winding numbers parametrizing the wrapping of the string around $S^1$, with respect to the two non-trivial cycles of the world-sheet torus. To obtain a Poincar\'e series representation of $\Gamma_{(1,1)}$, we separate out the $(\tilde{m},n)=(0,0)$ term and, in the remaining sum, we factor out the greatest common divisor (\emph{g.c.d.}) $N=(\tilde{m},n)$ of the non-vanishing windings. We can then express the windings as $\tilde{m}=Np$, $n=Nq$ with $(p,q)=1$ :
 \begin{align}
 	\Gamma_{(1,1)}(R) = R+2R\sum\limits_{N\geq 1}\biggr[\tfrac{1}{2}\sum\limits_{(p,q)=1}e^{-\frac{\pi (NR)^2}{\tau_2}|p+\tau q|^2}\biggr]~.
 \end{align}
The quantity inside the square brackets above is precisely a Poincar\'e series with seed $\varphi(\tau,\bar\tau) = \exp(-\frac{\pi(NR)^2}{\tau_2})$. This is easy to verify by noting that $\textrm{Im}(\gamma\cdot\tau_2)=\frac{\tau_2}{|p+\tau q|^2}$ for a matrix $\gamma=({\star\atop q}~{\star\atop p})\in \Gamma_\infty\backslash SL(2;\mathbb{Z})$. We then use this Poincar\'e series in order to unfold $\mathcal{F}$, as in (\ref{Unfold2}):
\begin{align}
	I = R\int\limits_{\mathcal{F}}d\mu~j(\tau) + 2R\sum\limits_{N\geq 1}\int\limits_0^\infty\frac{d\tau_2}{\tau_2^2}~e^{	-\frac{\pi(NR)^2}{\tau_2}}\int\limits_{0}^{1}d\tau_1~j(\tau)~.
\end{align}
Since $j(\tau)$ is defined with vanishing constant term in its (Fourier) $q$-expansion, the sum in the \emph{r.h.s.} vanishes due to the $\tau_1$-integration (level matching) and the integral (\ref{SimpleExample}) is simply given by the first integral in the \emph{r.h.s.} of the expression above. Being an integral of a holomorphic function (Case I of the previous section), this integral can be readily evaluated using Stokes' theorem, or using the techniques that we will review in the upcoming sections, $\int_{\mathcal{F}}d\mu~j(\tau)=-8\pi$ and, hence, we arrive at the following result:
\begin{align}\label{Unfold3}
	\int\limits_{\mathcal{F}}d\mu~ \Gamma_{(1,1)}(R)\,j(\tau) = -8\pi R~.
\end{align}
Now we encounter one of the major deficiencies of the traditional unfolding of $\mathcal{F}$ against the lattice. Namely, the result is \emph{not} invariant under $T$-duality, even though the lattice $\Gamma_{(1,1)}(R)$ in the \emph{l.h.s.} is  invariant under $R\rightarrow \frac{1}{R}$, as can be seen by Poisson resummation. The discrepancy is due to the loss of absolute convergence which, while being automatically ensured by the lattice in the UV, is only conditional in the IR ($\tau_2\rightarrow\infty$) and breaks down at precisely the T-self-dual radius $R=1$, due to the presence of `extra massless states'\footnote{The extra states becoming massless at $R=1$ are precisely saturated by the contribution of the unphysical `tachyon' pole of the $j$-function, responsible for the loss of exponential suppression in the IR.}. As a result, we are no longer allowed to exchange the order of integration and summation (\ref{Unfold2}) and the unfolding of $\mathcal{F}$ is no longer justified. The result (\ref{Unfold3}) is, in fact, only valid in the particular chamber of the moduli space, where $R>1$. In order to evaluate the integral for radii $R<1$, one should first double Poisson resum the lattice to its dual radius and then repeat the unfolding.

This simple example served to illustrate one of the most obvious deficiencies of the traditional unfolding approach. Indeed, using the orbit decomposition of the lattice in order to unfold $\mathcal{F}$ is only useful for extracting the large volume behaviour of the integral, with the loss of absolute convergence around extended symmetry points (\emph{i.e.} fixed points under T-duality) obscuring the behaviour of the amplitude around these regions. This is, in fact, a reflection of a much deeper drawback of the traditional unfolding method : using a Poincar\'e series representation of the lattice in order to unfold, does not yield the result in a manifestly T-duality invariant representation. The reason for this is that, unfolding against the lattice, inevitably starts with the lattice in its Lagrangian representation and relies on decomposing the winding sum into $SL(2;\mathbb{Z})$ orbits, with each orbit being used separately in order to unfold $\mathcal{F}$. However, the orbit decomposition of the winding sum leads to the loss of manifest T-duality invariance. Even though this may seem marginal in the simple example (involving a one-dimensional lattice) we presented above, as soon as one considers higher dimensional lattices, this problem becomes much more serious. For example, for the slightly more complicated integral involving a two-dimensional lattice and the almost holomorphic modular function $\hat{E}_2 E_4 E_6/\Delta$, one obtains:
\begin{align}\nonumber
	&\int\limits_{\mathcal F} d\mu\, \varGamma_{(2,2)} (T,U) \, \frac{\hat E_2 \, E_4 \, E_6}{\varDelta} \simeq  {\rm Re}\, \Biggl[ -24 \sum_{k>0} \left( 11\,  {\rm Li}_1 (e^{2\pi i k T}) - \frac{30}{\pi T_2\, U_2} {\mathcal P} (kT) \right)
\\ \nonumber
&- 24 \sum_{\ell >0} \left( 11 \, {\rm Li}_1 (e^{2\pi i \ell U}) - \frac{30}{\pi T_2\, U_2} {\mathcal P} (\ell U) \right)
 +\sum_{k>0, \ell >0} \left( \tilde c (k\ell ) \, {\rm Li}_1 (e^{2\pi i (kT+\ell U)}) - \frac{3\, c(k\ell ) }{\pi T_2\, U_2} {\mathcal P} (kT+\ell U) \right)
\\ \nonumber
&+ {\rm Li}_1 (e^{2\pi i (T_1 - U_1 +i |T_2 - U_2|)})
- \frac{3}{\pi T_2\, U_2} {\mathcal P} \left(T_1 - U_1 +i |T_2 - U_2|\right)\Biggr]
+\frac{60 \, \zeta (3)}{\pi^2 \, T_2 U_2} + 22  \log\left( \frac{8\pi e^{1-\gamma}}{\sqrt{27}}\, T_2 U_2 \right)
\\ \label{Complicated}
&+\left( \frac{4\pi}{3}\frac{U_2^2}{T_2}-\frac{22\pi}{3}U_2 - 4 \pi T_2 \right)\varTheta (T_2 - U_2)
+\left( \frac{4\pi}{3}\frac{T_2^2}{U_2}-\frac{22\pi}{3}T_2 - 4 \pi U_2 \right)\varTheta (U_2 - T_2)~,
\end{align}
where $\mathcal{P}(z)=\textrm{Im}(z)\,\textrm{Li}_2(e^{2\pi iz})+\frac{1}{2\pi}\textrm{Li}_3(e^{2\pi iz})$, $c,\tilde{c}$ are the Fourier coefficients of $E_4 E_6/\Delta$ and $E_2 E_4 E_6/\Delta$, respectively, and we are only displaying the IR finite part. Here, $\hat{E}_2,E_4,E_6$ are the weight 2 (almost holomorphic), weight 4 and weight 6 (holomorphic) Eisenstein series, respectively, and $\Delta=\eta^{24}$ is the weight 12 cusp form. As illustrated by  this example, even though the traditional unfolding method can be useful for extracting the asymptotic behaviour of amplitudes in the large volume limit, the results are generally \emph{local}, in the sense that they depend on  the region in moduli space around which one is unfolding\footnote{This is especially visible in  (\ref{Complicated}) from the Heaviside $\Theta$-functions in the last line.}. Even the task of merely checking the T-duality invariance of result (\ref{Complicated}) becomes a daunting task, whereas the singularity structure of the associated amplitude is fully obscured in this representation.

%%%%%%%

\section{New methods of unfolding}

The discussion above stresses the necessity for new methods of unfolding, able to overcome the limitations of the traditional unfolding method outlined in the previous section. In particular, one is ideally looking for a global\footnote{We use `global' here in order to stress the independence of the result from the region in moduli space around which one is unfolding.} representation of the result, which preserves the manifest T-duality symmetries of the lattice and which is able to capture the behaviour around the T-self-dual points. 
We will now proceed to briefly review two such techniques: the Rankin-Selberg-Zagier method and the unfolding against Niebur-Poincar\'e series, applicable to integrals of Cases II and III, respectively.

%%%%

\subsection{Integrals of Case II: the Rankin-Selberg-Zagier method}\label{sec:RSZ}

Integrals of Case II and, in general, modular integrals of a function of moderate (\emph{i.e.} non-exponential) growth at the cusp, can be treated using a technique known in the mathematics literature as the Rankin-Selberg-Zagier (RSZ) method \cite{MR656029,Angelantonj:2011br}. Here we will only point out the salient features applied to integrals of Case II. Notice first that the modular integral of a $d$-dimensional lattice $\Gamma_{(d,d)}$ has an IR divergence, since the integrand grows polynomially as $\tau_2^{d/2}$ at $\tau\rightarrow i\infty$ and needs to be regularized. The idea behind the RSZ method is to regularize the integral by truncating the fundamental domain to some (large) cutoff value $T$ and to deform the integrand by an insertion of the (completed) non-holomorphic Eisenstein series:
\begin{align}
	\int\limits_{\mathcal{F}}d\mu~\Gamma_{(d,d)}(G,B;\tau,\bar\tau) ~\longrightarrow~ \int\limits_{\mathcal{F}_T}d\mu~\Gamma_{(d,d)}(G,B;\tau,\bar\tau)\,E^\star(s;\tau)~.
\end{align}
Here, $\mathcal{F}_T=\{\tau\in\mathcal{F}~:~\tau_2<T\}$ is the truncated fundamental domain and the Eisenstein series is defined as:
\begin{align}
	E^\star(s;\tau)=\zeta^\star(2s)\sum\limits_{\gamma\in\Gamma_\infty\backslash SL(2;\mathbb{Z})}\bigr[\textrm{Im}(\gamma\cdot\tau)\bigr]^s = \tfrac{1}{2}\zeta^\star(2s)\sum\limits_{(c,d)=1}\frac{\tau_2^s}{|c\tau+d|^{2s}}~,
\end{align}
with $\zeta^\star(2s)=\pi^{-s}\Gamma(s)\zeta(2s)$ being the completed Riemann zeta function. Using the above Poincar\'e series representation for $E^\star$ we can unfold $\mathcal{F}_T$ for $\textrm{Re}(s)>1$:
\begin{align}\label{RSZ1}
	 \int\limits_{\mathcal{F}_T}d\mu~\Gamma_{(d,d)}(G,B;\tau,\bar\tau)\,E^\star(s;\tau)=\zeta^\star(2s)\int\limits_{S_T}d\mu~\tau_2^s\,\Gamma_{(d,d)}-\int\limits_{\mathcal{F}-\mathcal{F}_T}d\mu~\Gamma_{(d,d)}(E^\star(s;\tau)-\zeta^\star(2s)\tau_2^s)~,
\end{align}
 making sure to perform the appropriate subtraction of an infinite number of disks $S_{a/c}$ (with $a,c$ coprime integers, such that $c\geq 1$ and $a\in\mathbb{Z}_c$), corresponding to the images of the complement $\mathcal{F}-\mathcal{F}_T$ under $\gamma=\left({a\atop c}~{\star\atop\star}\right)\in SL(2;\mathbb{Z})$, which give rise to the second integral in the \emph{r.h.s} of (\ref{RSZ1}). It turns out that the latter becomes part of the definition of the renormalized integral. The next thing to notice is that $E^\star(s;\tau)$ is a meromorphic function with simple poles at $s=0,1$ with residue $\textrm{Res}_{s=1}E^\star(s;\tau)=\frac{1}{2}$ and, furthermore, $E^\star(s;\tau)=E^\star(1-s;\tau)$. Extracting the residue of both hand sides of (\ref{RSZ1}) at $s=1$ in order to obtain the desired integral and rearranging terms, one eventually obtains:
 \begin{align}\nonumber
 	\textrm{R.N.}\int\limits_{\mathcal{F}}d\mu~\Gamma_{(d,d)} & = 2\, \textrm{Res}_{s=1}\Biggr[\zeta^\star(2s)\int\limits_0^\infty d\tau_2~ \tau_2^{s+d/2-2}\int\limits_{0}^{1}d\tau_1~\sum\limits_{(m,n)\neq(0,0)}e^{-2\pi \tau_2\mathcal{M}^2}e^{2\pi i\tau_1 m^T n}\Biggr]\\ \label{RSZ2}
	&= 2\,\textrm{Res}_{s=1}\Biggr[ \zeta^\star(2s)\frac{\Gamma(s+\frac{d}{2}-1)}{\pi^{s+\frac{d}{2}-1}} \sum\limits_{(m,n)\neq(0,0)\atop m^T n=0}\frac{1}{[\mathcal{M}^2]^{s+\frac{d}{2}-1}}\Biggr]~,
 \end{align}
where the \emph{l.h.s.} is the renormalized integral\footnote{The definition of the renormalized integral and further details on the connection to other renormalization schemes can be found in \cite{Angelantonj:2011br}.} and the \emph{r.h.s.} is an integral over the half-infinite strip $S$ that can be easily performed. The constrained sum ($m^T n=m\cdot n=0$ projects onto $1/2$-BPS states) runs over the momentum and winding quantum numbers, $m,n$, respectively, excluding the origin $m_i=n^i=0$ and $\mathcal{M}^2$ is the (physical) BPS mass squared. It is related to the constrained Epstein zeta series constructed in \cite{Obers:1999um}. The result (\ref{RSZ2}) is manifestly invariant under T-duality, since the unfolding against the Eisenstein series does not depend on the point in moduli space around which we are unfolding. As a check, it is straightforward to use (\ref{RSZ2}) and the known Fourier expansion of $E^\star(s;\tau)$, in order to reproduce the standard results for the integrals of the $d=1$ and $d=2$ -dimensional lattices. Aside from manifestly preserving the T-duality symmetries, another upshot of the RSZ method is that it provides a natural modular invariant regularization. This should be contrasted with the traditional unfolding against a  $d=2$ lattice, where the degenerate orbit is IR divergent  and the regulator one introduces does not preserve modular invariance.

%%%%%%%

\subsection{Integrals of Case III: unfolding against Niebur-Poincar\'e series}

We now return to the generic integral (\ref{GenericBPSIntegral}), involving a (possibly asymmetric) lattice $\Gamma_{(d+k,d)}$ times a weak, almost holomorphic modular form $\Phi$ (elliptic genus) of weight $w=-k/2$ and with (at most) a simple pole in the $q$-expansion at the cusp. The presence of the pole in the $q$-expansion can be thought of as the contribution of the unphysical tachyon of the bosonic side of the Heterotic string. Due to the latter pole, the integrand grows exponentially at the cusp and the RSZ technique outlined in subsection \ref{sec:RSZ} is no longer applicable. A new method  \cite{Angelantonj:2012gw} is then required to treat integrals of Case III.

The main idea of \cite{Angelantonj:2012gw} is to construct a Poincar\'e series representation for the elliptic genus $\Phi$ itself and use it in order to unfold $\mathcal{F}$. The elliptic genus $\Phi$ can be expanded in the generators $\{\hat{E}_2,E_4,E_6,\Delta^{-1}\}$   of the graded ring of  weak almost holomorphic modular forms of weight $w$ and with a simple pole in $q$ at the cusp as:
\begin{align}
	\Phi(\tau) = \sum\limits_{2m+4n+6r=12+w\atop m,n,r\geq 0}{c_{mnr}~ \frac{\hat{E}_2^{m} E_4^{n} E_6^{r}}{\Delta}}~,
\end{align}
where $c_{mnr}$ are appropriate coefficients. The problem of determining the correct seed constructing the Poincar\'e series representation of $\Phi$ is a highly non-trivial one and we will not endeavor to present here the full details. Rather, it will suffice to mention only the guiding principles behind the construction. First, one notices that the hyperbolic Laplacian $\Delta_w=2\tau_2^2\partial_{\bar\tau}(\partial_\tau-\frac{iw}{2\tau_2})$ acts as a Casimir operator in the space of modular forms $\Phi$ of weight $w$ and the latter can be organized into appropriate linear combinations of its eigenmodes. The main idea is to construct a Poincar\'e series whose seed is already an eigenmode of $\Delta_w$. For generality, one imposes a pole of order $\kappa$ at the cusp $\Phi\sim q^{-\kappa}+\ldots$ and we are interested in constructing Poincar\'e series that are absolutely convergent for $w\leq 0$, so as to justify the unfolding. These conditions essentially lead to the choice of seed:
\begin{align}
	\varphi(\tau,\bar\tau)=\mathcal{M}_{s,w}(-\kappa\tau_2)\,e^{-2\pi i\kappa\tau_1}~,
\end{align}
where $\mathcal{M}_{s,w}(y)=|4\pi y|^{-w/2}\,M_{\frac{w}{2}\textrm{sgn}(y),s-\frac{1}{2}}(4\pi|y|)$ and $M$ is the Whittaker $M$-function. Summing over its images under $\Gamma_\infty\backslash SL(2;\mathbb{Z})$, this seed generates a Poincar\'e series known in the mathematics literature as the Niebur-Poincar\'e (NP) series \cite{0288.10010,0543.10020,1004.11021,BruinierOno,1154.11015,OnoMockDelta}:
\begin{align}\nonumber
	\mathcal{F}(s,\kappa,w)=\tfrac{1}{2}\sum\limits_{(c,d)=1}{(c\tau+d)^{-w}~\mathcal{M}_{s,w}\left(-\frac{\kappa\tau_2}{|c\tau+d|^2}\right)~\exp\left\{-2\pi i\kappa\left(\frac{a}{c}-\frac{c\tau_1+d}{c|c\tau+d|^2}\right)\right\}}~,
\end{align}
where the sum is over coprime integers $c,d$ and the integers $a,b$ are some solution of $ad-bc=1$. The NP series converges absolutely for $\textrm{Re}(s)>1$ and for $\kappa>0$, it indeed reproduces the desired pole of order $\kappa$ in the $q$-expansion at the cusp. By construction, it is an eigenmode of $\Delta_w$ with eigenvalue $-\frac{s(1-s)}{2}-\frac{w(w+2)}{8}$. Its spectrum can be obtained by studying its Fourier expansion and with the help of the modular derivatives $D_w=\frac{i}{\pi}(\partial_\tau-\frac{iw}{2\tau_2})$, $\bar{D}_w=-i\pi\,\tau_2^2\partial_{\bar\tau}$, acting as raising and lowering operators of the modular weight by units of $2$, respectively.

The NP series $\mathcal{F}(s,\kappa,w)$ transforms, by construction, in the same way as an holomorphic modular form. However, a careful study of its Fourier expansion reveals that it generically also contains a non-holomorphic part. What kind of modular objects do NP series represent ? In fact, the weak almost holomorphic modular forms we are interested in are eigenmodes of $\Delta_w$ with eigenvalue $-\frac{w}{2}$ and the NP series has the exact same eigenvalue\footnote{To be precise, the same eigenvalue also appears for NP series with $s=\frac{w}{2}$, but this lies outside the range of convergence for $w<0$, so that we can safely ignore it for the present discussion.} for $s=1-\frac{w}{2}$. It turns out, in general, that NP series $\mathcal{F}(1-\frac{w}{2},\kappa,w)$ are not (weak, almost) holomorphic modular forms but, rather, weak harmonic Maass forms. These are objects transforming like (weak, almost) holomorphic modular forms, but which are the sum of an holomorphic `Mock modular' part plus an infinite tower of negative frequency modes, explicitly breaking holomorphicity, called the `Shadow'. Hence, even though the holomorphic (Mock) part has an anomalous behaviour under modular transformations, this modular anomaly is precisely cancelled by the non-holomorphic (Shadow) part, which provides the modular completion. How then can we hope to obtain Niebur-Poincar\'e series representations of weak almost holomorphic modular forms $\Phi$? The answer is that, by taking appropriate linear combinations of NP series  $\mathcal{F}(1-\frac{w}{2},\kappa,w)$ with definite coefficients matching the principal part in the $q$-expansion of the modular form $\Phi$ we wish to represent, the Shadows cancel each other and the resulting linear combination precisely represents the given weak, holomorphic modular form. Weak almost holomorphic modular forms can then be formed out of similar --uniquely determined-- linear combinations of NP series with $s=1-\frac{w}{2}+n$.

Since all weak, almost holomorphic modular forms $\Phi$ can be uniquely expressed as linear combinations of absolute convergent Niebur-Poincar\'e series, we can effectively reduce the problem of evaluating the generic integral (\ref{GenericBPSIntegral}) into calculating the integral of the lattice $\Gamma_{(d+k,d)}$ times $\mathcal{F}(s,\kappa,-\frac{k}{2})$. The fundamental domain can now be unfolded using the Niebur-Poincar\'e series and the result is given in terms of a BPS sum:
\begin{align}\nonumber
	R.N.\int\limits_{\mathcal{F}}{d\mu~\Gamma_{(d+k,d)}~\,\mathcal{F}(s,\kappa,-\tfrac{k}{2})} &= \lim\limits_{T\rightarrow\infty}\left[\int\limits_{\mathcal{F}_T}{d\mu~\Gamma_{(d+k,d)}~\mathcal{F}(s,\kappa,-\tfrac{k}{2})}+f_0(s)\frac{T^{\frac{d}{2}+\frac{k}{4}-s}}{s-\frac{d}{2}-\frac{k}{4}}\right]\\ \label{BPSintegral}
 & = \int\limits_0^{\infty}{d\tau_2~\tau_2^{d/2-2}~\mathcal{M}_{s,-\frac{k}{2}}(-\kappa\tau_2)~\sum\limits_{\textrm{BPS}}~e^{-\pi\tau_2\,(P_L^2+P_R^2)/2}}~.
\end{align}
The first line provides the natural, modular invariant definition\footnote{This definition of the renormalized integral is valid for $\textrm{Re}(s)>\frac{d}{2}+\frac{k}{4}$ and can be extended, by meromorphic continuation, to the whole $s$-plane, except for the pole $s=\frac{d}{2}+\frac{k}{4}$, which requires a slightly modified subtraction of a logarithmic divergence (for details, see \cite{Angelantonj:2012gw}).} of the renormalized integral in terms of the appropriate cutoff-dependent subtraction, with $f_0(s)$ being the zero-frequency mode in the Fourier expansion of $\mathcal{F}(s,\kappa,w)$. The second line yields the result in terms of a strip integral, with the $\tau_1$-integration imposing the BPS constraint $P_L^2-P_R^2=\kappa$, whereas the $\tau_2$-integral casts the BPS-contribution in its Schwinger representation. The $\tau_2$-integral can be explicitly performed to yield the BPS sum:
\begin{align}\label{NPresult}
	I=(4\pi\kappa)^{1-\frac{d}{2}}\,\Gamma(s+\tfrac{d}{2}+\tfrac{k}{4}-1)\sum\limits_{\textrm{BPS}}{ {}_2 F_1\left(s-\frac{k}{4},s+\frac{d}{2}+\frac{k}{4}-1;2s;\frac{4\kappa}{P_L^2}\right)\,\left(\frac{P_L^2}{4\kappa}\right)^{1-s-\frac{d}{2}-\frac{k}{4}} }~,
\end{align}
which can be proven to converge absolutely for $\textrm{Re}(s)>\frac{d}{2}+\frac{k}{4}$ and can be meromorphically continued to the full $s$-plane with the exception of a simple pole at $s=\frac{d}{2}+\frac{k}{4}$ which requires an additional subtraction. The result (\ref{NPresult}) is manifestly invariant under T-duality and chamber independent, providing a global representation of the result, valid at any point in moduli space. Therefore, it can be applied to study the behaviour of amplitudes around points of extended symmetry, where the traditional method of unfolding breaks down. In fact, for all cases of interest to string theory applications, it is possible to re-express the hypergeometric function in (\ref{NPresult})  in terms of elementary functions. The general expression is given in \cite{Angelantonj:2012gw} and renders the singularity structure of the integral crystal clear. One finds that that for odd-dimensional lattices, the integral (\ref{BPSintegral}) always develops conical singularities, whereas for $d\geq 3$ real singularities also appear. On the other hand, only real singularities appear in even dimensions, including power-like singularities for $d\geq 4$ and logarithmic ones for $d\leq 2s+w$. Furthermore, in the absence of Wilson lines, one may prove that the universal singularity behaviour in $d=2$ dimensions has the form:
\begin{align}
	\int\limits_{\mathcal{F}}d\mu~\Gamma_{(2,2)}\mathcal{F}(1+n,1,0) \simeq -\frac{(2n+1)!}{n!}~\log|j(T)-j(U)|^4~.
\end{align}
Furthermore, the explicit expression for the BPS sum (\ref{NPresult}) in \cite{Angelantonj:2012gw} can be used to prove, in a chamber independent fashion, the absence of singularities in gauge thresholds involving elliptic genera where the unphysical tachyon pole cancels out\footnote{Note that, using the traditional unfolding method of Section \ref{sec:TraditionalUnfolding}, such a proof is \emph{a priori} not possible, even if one asymptotically approaches the boundary separating different chambers of moduli space around an enhancement point.}. In the last section, we will illustrate the power of our new method by applying it to specific examples.

%%%%%

\section{Examples}

We will first start with integrals of Case III involving a one-dimensional lattice:
\begin{align}\label{1dintegral}
	\int\limits_{\mathcal{F}}{d\mu~\Gamma_{(1,1)}(R)~\mathcal{F}(1+n,1,0)}=2^{2+2n}\,\sqrt{\pi}\,\Gamma(n+\tfrac{1}{2})\,\left(\,R^{1+2n}+\frac{1}{R^{1+2n}}-\left|R^{1+2n}-\frac{1}{R^{1+2n}}\right|\,\right)~,
\end{align}
for any integer $n\geq 0$. For example, for $n=0$, one has $\mathcal{F}(1,1,0)=j(\tau)+24$. With the help of  the elementary result $\int_{\mathcal{F}}d\mu~\Gamma_{(1,1)}(R)=\frac{\pi}{3}(R+\frac{1}{R})$, which can be derived \emph{e.g.} using the RSZ method, we immediately derive:
\begin{align}
	\int\limits_{\mathcal{F}}d\mu~\Gamma_{(1,1)}(R)j(\tau) = -4\pi \left(\,R+\frac{1}{R}+\left|R-\frac{1}{R}\right|\,\right)~.
\end{align}
The result is manifestly invariant under T-duality and holds for any radius. It should be contrasted with (\ref{Unfold3}), which is only valid in the $R>1$ chamber and fails to display the T-duality symmetries of the $S^1$-lattice. On the other hand, the conical singularity appears naturally within our formalism. Another  example is the one-dimensional analogue of gravitational thresholds in $E_8\times E_8$ Heterotic string theory compactified on $K3\times T^2$ :
\begin{align}
	\int\limits_{\mathcal{F}}d\mu~\Gamma_{(1,1)}(R)\frac{\hat{E}_2 E_4 E_6}{\Delta} = 8\pi(R^3+R^{-3}-|R^3-R^{-3}|)-68\pi(R+R^{-1})+20\pi|R-R^{-1}|~,
\end{align}
which can be easily obtained using the Niebur-Poincar\'e series expansion  $\hat{E}_2 E_4 E_6/\Delta = \mathcal{F}(2,1,0)-5\mathcal{F}(1,1,0)-144$.

Let us now consider examples of $\mathcal{N}=2$ thresholds of the $E_8\times E_8$ Heterotic string compactified on $K3\times T^2$, where we realize $K3$ as a $T^4/\mathbb{Z}_2$ orbifold. The gauge thresholds for the $E_8$ and $E_7$ factors are given by the BPS-sums:
\begin{align*}
	&\Delta_{E_8}=-\frac{1}{12}\int\limits_{\mathcal{F}}{d\mu~\Gamma_{(2,2)}(T,U)~\frac{\hat{E_2}E_4 E_6-E_6^2}{\Delta}}=\sum\limits_{\textrm{BPS}}{\Bigr[1+\frac{P_R^2}{4}\,\log\left(\frac{P_R^2}{P_L^2}\right)\Bigr]}+72\,\log\Bigr(T_2 U_2|\eta(T)\eta(U)|^4\Bigr)+\textrm{cst}~,\\
	&\Delta_{E_7}=-\frac{1}{12}\int\limits_{\mathcal{F}}{d\mu~\Gamma_{(2,2)}(T,U)~\frac{\hat{E_2}E_4 E_6-E_4^3}{\Delta}}=\sum\limits_{\textrm{BPS}}{\Bigr[1+\frac{P_R^2}{4}\,\log\left(\frac{P_R^2}{P_L^2}\right)\Bigr]}-72\,\log\Bigr(T_2 U_2|\eta(T)\eta(U)|^4\Bigr)+\textrm{cst}~.
\end{align*}
The case of non-trivial Wilson lines can also be easily treated. If we Higgs the $E_8$ group factor down to its Coulomb branch, the $E_7$ threshold becomes:
\begin{align*}
	\Delta_{E_7}=-\frac{1}{12}\int\limits_{\mathcal{F}}{d\mu~\Gamma_{(2,10)}~\frac{\hat{E_2}E_6-E_4^2}{\Delta}}=\sum\limits_{\textrm{BPS}}{\Biggr[1+\frac{P_R^2}{4}\,\log\left(\frac{P_R^2}{P_L^2}\right)-\frac{2}{P_L^2}-\frac{8}{3\,P_L^4}-\frac{16}{3\,P_L^6}-\frac{64}{5\,P_L^8}\Biggr]}~,
\end{align*}
where the left- and right- moving momenta $P_{L,R}$ now also depend on the Wilson lines $Y^a_I$ and the BPS constraint now involves also  the $U(1)$ charge vectors $Q$ in the Cartan of $E_8$, $m^T n+\frac{1}{2}Q^T Q=1$. It is straightforward to verify that the amplitudes are regular\footnote{This is physically expected, since the unphysical tachyon of the Heterotic string is chargeless with respect to $E_7$ and $E_8$.} at any point in moduli space and valid in any chamber.

A further application of our new methods of unfolding concerns the treatment of integrals involving insertions of $P_{L,R}$ lattice momenta of the generic form:
\begin{align}
	\int\limits_{\mathcal{F}}{d\mu~\Biggr[\tau_2^{-\lambda/2}\,\sum\limits_{P_L,P_R}{\rho(P_L\sqrt{\tau_2},P_R\sqrt{\tau_2})~q^{\frac{1}{4}P_L^2}\,\bar{q}^{\frac{1}{4}P_R^2}}\Biggr]~\Phi(\tau)}~.
\end{align} 
For consistency, the quantity in the brackets must be a modular form of weight $\lambda+d+\frac{k}{2}$. The general conditions on the function $\rho(x,y)$ for this to take place can be found in \cite{Vigneras,Angelantonj:2012gw}. Provided they are satisfied, the integrand is modular invariant and $-w=\lambda+d+\frac{k}{2}$. One may then expand the elliptic genus $\Phi$ in terms of Niebur-Poincar\'e series and unfold against each of them to obtain the corresponding BPS sum:
\begin{align*}
	\int\limits_{\mathcal{F}}&{d\mu~\tau_2^{-\lambda/2}\,\sum\limits_{P_L,P_R}{\rho(P_L\sqrt{\tau_2},P_R\sqrt{\tau_2})~q^{\frac{1}{4}P_L^2}\,\bar{q}^{\frac{1}{4}P_R^2}}~\mathcal{F}(s,\kappa,w)} \\
 &= (4\pi\kappa)^{1+\lambda/2}\int\limits_{0}^{\infty}{dt~t^{2+\frac{2d+k}{4}-2}{}_1 F_1\left(s-\frac{2\lambda+2d+k}{4};2s;t\right)~\rho\left(P_L\sqrt{\frac{t}{4\pi\kappa}},P_R\sqrt{\frac{t}{4\pi\kappa}}\right)~\sum\limits_{\textrm{BPS}}~e^{-t P_L^2/4\kappa}}~.
\end{align*}

In order to demonstrate the power of our methods, we will present one final example involving an `exotic' integral that does not even contain moduli dependence:
\begin{align}\label{exoticIntegral}
	\int\limits_{\mathcal{F}}{d\mu~\bigr(\sqrt{\tau_2}\eta\bar\eta\bigr)^3~\frac{\hat{E}_2^2\,E_8-2\,\hat{E}_2\,E_{10}}{\Delta} } = -20\sqrt{2}~.
\end{align}
This integral is not only interesting as a mathematical exercise but, in fact, appears in \cite{Carlevaro:2012rz} as a threshold contribution to certain non-compact Heterotic constructions on ALE spaces in the presence of NS5 branes. This is an example where the traditional orbit method cannot even be applied. Expanding the weak almost holomorphic modular form into Niebur-Poincar\'e series:
\begin{align}
	\frac{\hat{E}_2^2 E_8-2\hat{E}_2 E_{10}}{\Delta}=\tfrac{1}{5}\mathcal{F}(3,1,0)-6\mathcal{F}(2,1,0)+23j+984~,
\end{align}
and using (\ref{1dintegral}), it is straightforward to arrive at the explicit numerical value $-20\sqrt{2}$ in (\ref{exoticIntegral}). This simple example illustrates that our novel methods can also be applied to integrals that do not even contain the Narain lattice.

%%%%

\section{Conclusions}

In this short review we attempted to briefly portray some of the aspects of our novel approach to evaluating one-loop BPS-saturated amplitudes in string theory. We discussed how the traditional unfolding method, using the orbit decomposition of the lattice, generically fails to preserve the manifest T-duality symmetries of the theory. Our novel proposal was to exploit the fact that any weak, almost holomorphic modular form (such as the modified elliptic genus) can be uniquely represented as a linear combination of absolutely convergent Niebur-Poincar\'e series and we can use these to unfold the fundamental domain $\mathcal{F}$. BPS-saturated one-loop string amplitudes are then naturally expressed as sums over the perturbative BPS states in a manifestly T-duality invariant fashion. Within this new framework, the singularity structure of the amplitudes becomes crystal clear and the results in this representation are valid at any point in moduli space (chamber independent). The incorporation of non-trivial Wilson lines and lattice momentum insertions is also achieved in a simple manner, as illustrated in several examples. Finally, these methods successfully apply to cases of `exotic' integrals, that may not even involve moduli dependence.

We end this short review with a few comments concerning integrals of Case IV, namely, the class of integrals where the integrand function $\mathcal{A}$ is manifestly non-holomorphic. Examples of this class in string theory include, \emph{e.g.} the non-trivial 1-loop corrections to the effective potential (vacuum energy) of  Type II and Heterotic vacua with spontaneously broken supersymmetry, or the technically similar case of the free energy of string theories at finite temperature. We would like to note that, traditionally, the behaviour of the 1-loop effective potential around points of extended symmetry has been notoriously hard to study. In fact, understanding these properties is highly related to some of the long-standing puzzles plaguing string thermodynamics and string cosmology, such as the resolution of the Hagedorn phase transition and the initial singularity problem. Even though, with our present machinery, attacking  the fully non-holomorphic integrals of Case IV seems to be out of reach (aside from certain notable exceptions), it is still hoped that future progress in this direction may provide the  tools necessary to study such integrals as well.

\section*{Acknowledgements}
It is a pleasure to thank my collaborators C. Angelantonj and B. Pioline for a very enjoyable collaboration and the organizers of the \emph{Corfu Summer Institute 2012} and the \emph{XVIII European Workshop on String Theory}  for giving me the opportunity to present this work.

\end{document}